\documentclass[conference]{IEEEtran}
\IEEEoverridecommandlockouts
\usepackage{amsmath,amssymb,amsfonts}
\usepackage{algorithmic}
\usepackage{graphicx}
\usepackage{textcomp}
\usepackage[table,xcdraw]{xcolor}
\usepackage{balance}
\usepackage{soul}
\usepackage{booktabs}
\usepackage{multirow}
\usepackage{float}
\usepackage{url}
\usepackage{tcolorbox}
\usepackage{tabularx} 
\usepackage[noadjust]{cite}
\usepackage[hidelinks,bookmarks=true,pagebackref=true]{hyperref}
\usepackage{xurl}
\usepackage{enumitem}
\setlist[itemize,enumerate]{noitemsep, topsep=0pt, leftmargin=1.0em}
\usepackage[utf8]{inputenc}
\usepackage{newunicodechar,graphicx}
\DeclareRobustCommand{\okina}{%
  \raisebox{\dimexpr\fontcharht\font`A-\height}{%
    \scalebox{0.8}{`}%
  }%
}
\newunicodechar{ʻ}{\okina}
\usepackage{listings} 
\usepackage{xcolor}   

\lstdefinestyle{pythonstyle}{
  language=Python,
  basicstyle=\ttfamily\small,    
  keywordstyle=\bfseries\color{blue},   
  commentstyle=\itshape\color{green!50!black}, 
  stringstyle=\color{red},   
  frame=single,               
  numbers=none,              
  numberstyle=\tiny\color{gray}, 
  stepnumber=1,              
  tabsize=4,                 
  showstringspaces=false,    
  breaklines=true,           
  showspaces=false,
  captionpos=b,
  rulecolor=\color{black}
}
\usepackage{titlesec}
\titleformat{\subsubsection}
  [runin] 
  {\normalfont\bfseries\itshape} 
  {\thesubsubsection} 
  {1em} 
  {} 
  [:\,] 
\usepackage{background}
\backgroundsetup{
  position=current page.east,
  angle=-90,
  nodeanchor=east,
  vshift=-4mm,
  opacity=0.5,
  scale=3,
  contents=PREPRINT
}

\newcommand{\RQA}{\textbf{RQ1}: What are the backgrounds and extent of programming experiences of research scientists?}

\newcommand{\RQB}{\textbf{RQ2}: What are the practices, challenges, and perceptions regarding code understandability in scientific programming?}

\newcommand{\RQC}{\textbf{RQ3}: What is the importance of identifier names in scientific software readability, and what challenges arise in this area?}

\begin{document}

\title{\huge Exploring Code Comprehension in Scientific Programming: Preliminary Insights from Research Scientists}

\author{
\IEEEauthorblockN{Alyssia Chen, Carol Wong}
\IEEEauthorblockA{\textit{University of Hawaiʻi at Mānoa} \\
Honolulu, Hawaiʻi, USA \\
abc8@hawaii.edu, carolw8@hawaii.edu}
\and
\IEEEauthorblockN{Bonita Sharif}
\IEEEauthorblockA{\textit{University of Nebraska - Lincoln} \\
Lincoln, Nebraska, USA \\
bsharif@unl.edu}
\and
\IEEEauthorblockN{Anthony Peruma}
\IEEEauthorblockA{\textit{University of Hawaiʻi at Mānoa} \\
Honolulu, Hawaiʻi, USA \\
peruma@hawaii.edu}
}

\maketitle

\begin{abstract}
Scientific software—defined as computer programs, scripts, or code used in scientific research, data analysis, modeling, or simulation—has become central to modern research. However, there is limited research on the readability and understandability of scientific code, both of which are vital for effective collaboration and reproducibility in scientific research. This study surveys 57 research scientists from various disciplines to explore their programming backgrounds, practices, and the challenges they face regarding code readability. Our findings reveal that most participants learn programming through self-study or on-the-job training, with 57.9\% lacking formal instruction in writing readable code. Scientists mainly use Python and R, relying on comments and documentation for readability. While most consider code readability essential for scientific reproducibility, they often face issues with inadequate documentation and poor naming conventions, with challenges including cryptic names and inconsistent conventions. Our findings also show low adoption of code quality tools and a trend towards utilizing large language models to improve code quality. These findings offer practical insights into enhancing coding practices and supporting sustainable development in scientific software.
\end{abstract}

\section{Introduction}
\label{Section:Introduction}
Scientific software, which includes computer programs, scripts, algorithms, code, or models, plays a vital role in advancing scientific research \cite{Arvanitou2021}. Advancements in tools, Integrated Development Environments (IDEs), and software frameworks and libraries have made the development of scientific software more accessible to researchers across various fields \cite{vaughan2023python,Arvanitou2021}. With scientists spending approximately 30\% to 35\% of their time developing software to support their research (\cite{Hannay2009,Prabhu2011}), the quality of scientific code has become vital for research sustainability, reproducibility, and collaboration.

These challenges connect directly to program comprehension, a core practice in software development (\cite{Mayrhauser1995}), where programmers can spend 58\% of their time dedicated to analyzing code to understand its functionality for tasks such as bug fixing or feature changes \cite{Xia2018}. This importance of code quality in enhancing comprehension has been extensively studied by researchers, focusing on various factors such as code structure, the semantics and structure of identifier names, and comments, among other characteristics \cite{Sergeyuk2024, Arnaoudova2016,Li2020Renamings,NewmanCoRR2020,LawrieICPC2006,Kaur2020,ParkEMSE24}. However, these studies have primarily focused on general-purpose or consumer-oriented software systems \cite{Wyrich2023}.

Even though scientific programs are written in the same programming languages as other software systems (\cite{Prabhu2011,Rule2018}), they present distinct comprehension challenges. Scientific programs are written by scientists who prioritize domain knowledge over established software development practices \cite{Segal2008}. Further, development environments like Jupyter Notebooks, while making programming more accessible through integrated live code, narrative text, and visualizations, introduce their own challenges: non-linear execution flows, hidden states, lack of modularity, contains redundant code, and organizational issues \cite{Head2019,Adams2023,Pimentel2021,Chattopadhyay2020,Rule2018}. Moreover, empirical studies show that Jupyter Notebooks tend to have more quality issues than traditional Python scripts, with higher rates of PEP8 violations, more stylistic problems,  higher coupling, and often suffer from reproducibility challenges \cite{Wang2020,grotov2022large,Adams2023,PimentelMSR2019}. 

\subsection{Study Goal}
\label{Section:Goal}
While these prior studies demonstrate that scientific programs are not immune from code comprehension challenges, there is limited knowledge in our understanding of how scientific programmers themselves perceive and address these issues
. \textit{To bridge this gap, in this study, we survey research scientists across various disciplines, aiming to understand their perspectives on code comprehension, their current practices, and the challenges they face in maintaining readable and understandable scientific code.} We envision that this early-stage research will serve as a foundation for more comprehensive studies into the factors influencing code comprehension in scientific programming, potentially leading to the development of new tools and practices tailored for scientific programmers.

\subsection{Contributions}
\label{Section:Contributions}
The main contributions of this work are:
\begin{itemize}
    \item Insights from scientific programmers conducting research across multiple disciplines regarding their approach to code comprehension in practice. 
    \item Empirical evidence identifying the key challenges in scientific code comprehension, including insufficient documentation, poor identifier naming, and limited use of code quality tools, with a trend towards the adoption of AI-based code generation tools.
    \item Findings that serve as a foundation for future research in scientific code comprehension by identifying multiple research directions that aim to develop targeted solutions tailored to the specific needs of scientific programmers and improve code quality in research software.
\end{itemize}

\section{Study Design}
\label{Section:method}
This section describes our methodology.

\subsection{Survey Design}
We used the Qualtrics platform \cite{Qualtrics} to design a 20-question survey aimed at collecting data on participants' experiences and perceptions of code readability in scientific programming. The questions were developed based on the objectives of our study and a review of relevant literature. Additionally, following best practice \cite{linaaker2015guidelines}, we conducted a pilot run with three research scientists, adjusting the questionnaire based on their feedback. The survey included a mix of single-choice, multiple-choice, and free-text questions. Due to space constraints, the complete survey instrument is not included in this paper. However, to ensure reproducibility and transparency, all relevant materials are available in our artifact package at \cite{ArtifactPackage}.

\subsection{Survey Participants}
Although this is an exploratory study, we aimed to obtain a diverse set of participants. Hence, we contacted researchers in multiple departments within the University of Hawaiʻi at Mānoa via email, some of whom forwarded our invitation or referred us to their colleagues or lab members. This combination of convenience and snowball sampling allowed us to identify qualified individuals \cite{Baltes2022}. Participation was voluntary and anonymous, and no compensation was provided. We contacted 600 individuals and received 112 responses. To ensure consistency in our analysis of results, we only considered participants who answered all mandatory questions, resulting in 57 valid responses for analysis.

\subsection{Data Analysis}
We analyzed the data using quantitative and qualitative techniques \cite{Wagner2020}. 
For the quantitative analysis, we used descriptive statistics, particularly for the single- and multi-choice questions. For the open-ended responses, we conducted a systematic thematic analysis. Three authors independently reviewed the data to identify and categorize key concepts, resolving any coding disagreements through consensus meetings.


\section{Results}
\label{Section:results}
We received 57 completed responses (all mandatory questions answered). Due to space constraints, in some parts of the writeup, we only report on the frequent observations. Our artifact package (\cite{ArtifactPackage}) contains the complete dataset and the thematic coding of free-text responses. 

\subsection*{\RQA}
\textbf{Motivation.} This RQ explores the educational qualifications and programming experience of research scientists. This knowledge provides insight into how scientists acquire programming skills and the extent of their programming expertise. The findings can help identify gaps in programming education and explain their software development practices and the challenges they face in code understandability and maintainability. 

\subsubsection*{General Background} Most participants (51, or 89.47\%) hold PhD degrees, followed by five with Master’s degrees and one with a Bachelor’s degree. We identify 35 unique majors among the 57 participants, reflecting a diverse range of educational specializations. Specifically, 11 participants have an economics background, six come from computer science, and three have a background in physics. The most common professions include professors, with 41 participants (71.93\%), followed by six graduate students and four research scientists. When asked about their primary domains of work, the top three responses are 11 participants (12.50\%) working in Economics and 10 each in Computer Science and Environmental Science.

\subsubsection*{Programming Experience} Python and R are the two most popular programming languages participants use for writing scientific programs, with 31 (27.93\%) reporting Python and 29 (26.13\%) reporting R. In contrast, languages such as C\# and Java are reported by four and two participants, respectively. Since this is a multiple-choice question, the average number of languages reported by participants is 1.94, with Python and R being the common combination, occurring 14 times (18.42\%). Further, participants report using statistical software such as Stata and SAS. Regarding preferred coding environments, most participants (46 or 43.81\%) favor traditional IDEs (e.g., PyCharm, Visual Studio, etc.), followed by text editors (e.g., Sublime Text, Atom, Notepad++, etc.), which 23 participants utilize. Jupyter Notebooks and similar notebook interfaces/services are reported by 15 participants. On average, participants use 1.84 environments, with the most common pair being text editors and traditional IDEs, reported 14 times.

When asked where they initially learned to program, the top three answers are self-taught through online articles and blogs (31 responses or 22.30\%), on-the-job training or learning (29 or 20.86\%), and formal computer science courses at a university (23 or 16.55\%). On average, participants select 2.44 options for this multiple-choice question, with the most common pair being on-the-job training or learning and being self-taught through online articles and blogs, which occurs 19 times. Moving on, 33 participants (57.9\%) report that they have not received any education or training in writing readable and maintainable code. Among those who do receive such education/training, an analysis of free-text responses shows that most participants learn through formal education, while others acquire their knowledge from on-the-job experience, workshops, or peers. The topics covered in their education/training include coding best practices, documentation, testing practices, and tools designed to enhance code quality.

\subsection*{\RQB}
\textbf{Motivation.} Code comprehension is an essential activity in software development and maintenance. This RQ identifies the specific practices scientists employ to write understandable code and the obstacles they encounter when comprehending scientific code. The findings provide insight into crucial areas to support scientists in writing understandable code, leading to improved reproducibility of scientific results. 

\subsubsection*{Practices} We asked participants, using a free-text question, what practices they frequently employ to improve code readability. An analysis of their responses shows that documentation practices, particularly code commenting, is a common approach (63.16\% of responses), while 15.79\% of participants report using multiple practices, including documentation, modular programming, and naming conventions. The remaining responses were distributed across specific focuses, such as using AI tools (e.g., ChatGPT) and refactoring. When surveyed about their use of automated code quality tools to improve code readability, 49.12\% of participants indicate they `Never' use them. Among those who do, many utilize AI and large language models (LLMs), particularly ChatGPT and Claude, to enhance code readability, accounting for 41.38\% of responses. Approximately 24\% mention using a combination of tools, including IDEs, LLMs, and online resources. The remaining responses focus on specific tools, with IDE features (e.g., refactoring and styling) used by 14\% of participants.

\subsubsection*{Challenges} Through a Likert question, 
54.55\% of participants report that they `Sometimes' encounter difficulties in understanding scientific software source code written by others, while 38.18\%, 5.45\%, and 1.82\% select `Often,' `Rarely,' and `Always,' respectively. Analyzing the challenges participants frequently encounter, we observe (as shown in Table \ref{Table:rq2-challenge}) that a lack of documentation in the form of code comments and project-level documentation (e.g., READMEs) are the top two challenges reported by 44 and 33 participants, respectively, closely followed by poor identifier naming and project structure, each with 31 instances. On average, participants select 4.77 challenges in this multi-choice question. Three challenges that frequently occur together (18 occurrences) are lack of project documentation, insufficient comments, and poor naming of methods/variables.

\begin{table}[t]
\centering
\caption{Top five frequent challenges comprehending scientific code.} 
\vspace{-2mm}
\label{Table:rq2-challenge}
\resizebox{\columnwidth}{!}{%
\begin{tabular}{@{}p{0.43\linewidth}rr@{}} 
\toprule
\multicolumn{1}{c}{\textbf{Challenge}} & \multicolumn{1}{c}{\textbf{Count}} & \multicolumn{1}{c}{\textbf{Percentage}} \\ \midrule
Lack of or insufficient comments                  & 44 & 16.18\% \\
Lack of documentation (e.g., no README)            & 33 & 12.13\% \\
Poor naming of methods/functions, variables, etc. & 31 & 11.40\% \\
Poor organization of project structure            & 31 & 11.40\% \\
Hardcoded values without explanation              & 24 & 8.82\%  \\ \bottomrule
\end{tabular}%
}\vspace{-2mm}
\end{table}

\subsubsection*{Perception on Reproducibility} All participants acknowledge the importance of code readability for ensuring reproducible scientific results, as indicated by a Likert-scale question. Specifically, 26 participants (45.16\%) find it `Very important,' while 22 participants (38.60\%) rate it as `Extremely important,' Five participants report it as `Slightly important,' and four consider it `Moderately important.' None of the participants select the option `Not important at all.'

\subsection*{\RQC}
\textbf{Motivation.} High-quality identifier names are crucial for understanding the correct behavior of code \cite{Schankin2018}. This RQ investigates scientists' views on identifier names and identifies the specific naming challenges they face with scientific code. The findings can guide the creation of automated tools to enhance naming practices in scientific coding.

\subsubsection*{Importance} Our 5-point Likert scale question regarding the importance of identifier names in making code readable shows that all participants recognize the importance of names. Specifically, 25 participants (43.86\%) consider them `Very important,' while 17 (29.82\%) view them as `Extremely important,' and 13 (22.81\%) rate them as `Moderately important.' Only two participants rate it as `Slightly important,' and none select `Not important at all.'

\subsubsection*{Challenges}
When asked about the frequency at which they experience misunderstanding or errors in the code due to identifier names, 32 (or 56\%) participants reported `Sometimes,' while eighteen, five, and two participants reported `Often,' `Rarely,' and `Never,' respectively. To gain more insight into their challenges, participants were presented with a multi-choice question regarding the common challenges they encounter with identifier names. As shown in Table \ref{Table:rq3-challenge}, the two most common issues were names that were too short or cryptic and names that followed inconsistent naming conventions, as reported by 40 (21.28\%) and 30 (15.96\%) participants, respectively. Participants chose, on average, 3.3 options for this question, with the most common two-pair combination being names that are too short or cryptic and the use of generic names. Finally, through a free-text question, participants were asked to provide examples of naming issues they have encountered. An analysis of their responses revealed issues, such as ambiguous and misleading names, using similar or same names for different purposes, overly short names (e.g., single characters), and inconsistencies in naming styles.

\begin{table}
\centering
\caption{Top five frequent challenges with identifier names.}
\vspace{-2mm}
\label{Table:rq3-challenge}
\resizebox{\columnwidth}{!}{%
\begin{tabular}{@{}p{0.5\linewidth}rr@{}} 
\toprule
\multicolumn{1}{c}{\textbf{Challenge}}                                                & \multicolumn{1}{c}{\textbf{Count}} & \multicolumn{1}{c}{\textbf{Percentage}} \\ \midrule
Names that are too   short or cryptic (e.g., single letters abbreviations or acronyms) & 40                                 & 21.28\%                                 \\
Inconsistent naming conventions and   styles         & 30 & 15.96\% \\
Names that don't accurately reflect their   purpose  & 29 & 15.43\% \\
Use of generic names (e.g., 'data', 'result', 'temp') & 29 & 15.43\% \\
Use of names that are too similar                    & 25 & 13.30\% \\ \bottomrule
\end{tabular}%
}
\end{table}

\section{Discussion}
\label{Section:discussion}
This preliminary study introduces a novel perspective on code comprehension in the context of scientific programming. While previous work has extensively focused on code comprehension in traditional software development, our study forms the foundation for examining the specific practices and challenges with scientific code comprehension. 

\subsection{Key Findings and Their Implications}

\noindent\textbf{\textit{Extending programming education beyond syntax mastery:}} Our findings from RQ 1 support previous research showing that many scientists lack formal education in computer science or software engineering \cite{Leroy2022,Santana2024}. Additionally, our results show that these scientists mainly develop their programming skills through self-study and practical experience, which aligns with earlier research \cite{Hannay2009}. However, our findings also offer additional insights showing that scientists typically lack an understanding of the best practices in writing maintenance-friendly code. These observations are important as they suggest potential risks to the long-term sustainability of research artifacts, which can include code misinterpretation, higher cognitive load, increased debugging time, and challenges in collaborative projects \cite{Soh2016,Fakhoury2018}. These findings emphasize the importance of expanding programming education beyond just mastering syntax to include important software engineering practices. Hence, it is essential for academic and research institutions to implement training programs that focus on key software engineering concepts for scientists, establish coding standards for research software, and provide researchers with centralized resources for supporting code maintenance.

\noindent\textbf{\textit{Need for focused program comprehension research on scientific programming:}} The study also highlights the scientific community's strong preference for Python and R over languages like Java and C\#, aligning with prior research \cite{Prabhu2011,Rule2018}. Further, through RQ2, we identified code readability as a critical factor in scientific software, especially for ensuring reproducibility. RQ3 highlights the importance of high-quality identifier names in achieving this readability. These findings are of particular importance as prior research on code comprehension has predominantly concentrated on Java programs, examining aspects such as readability models, linguistic antipatterns, and identifier naming \cite{Sergeyuk2024,Arnaoudova2016,Li2020Renamings}, showing that there is a clear need for targeted research that focuses on the unique aspects of scientific programming. Additionally, most research studies on code comprehension typically focus on code written in Java (\cite{Wyrich2023}), further highlighting the need for research pertaining to the specific characteristics of scientific programming languages like Python and R. Finally, it is important for studies involving human subjects to broaden their recruitment beyond the traditional pool of computer science students and include researchers from various fields.

\noindent\textbf{\textit{Need for better tool support and AI usage guidelines:}} Our RQ 2 findings reveal an interesting pattern in how scientific programmers approach code quality tools. While there is low adoption of traditional code quality tools (e.g., \cite{Rule2019}), AI-powered tools, especially LLMs, are popular. This low adoption of traditional tools may indicate either a lack of knowledge among scientists or the inadequacy of these tools in addressing the specific needs of scientific programming and warrants further investigation. On the other hand, the shift toward LLMs mirrors trends in software engineering, where such AI-based tools are enhancing developer productivity \cite{Sauvola2024,Liang2024}. However, studies have shown that LLM-generated code can contain quality issues \cite{Dantas2023,Siddiq2024}. This is particularly problematic in scientific programming, where many practitioners lack formal software engineering training. Hence, it is essential that they are equipped with the skills to critically evaluate AI-generated code \cite{Liu2024}, especially in a scientific context.

\noindent\textbf{\textit{The documentation paradox:}} Our findings indicate that while documentation is the most commonly reported approach to improving code understandability, paradoxically, our findings also show that the lack of or insufficient comments and project documentation emerge as the top challenges in code comprehension. This contradiction presents multiple research opportunities, such as investigating what constitutes ``sufficient'' documentation across various scientific fields, understanding the gap between perceived and actual documentation practices, and examining its impact on reproducibility. Such studies can lead to improved documentation guidelines and document-generation tools for scientific code \cite{Rule2019,Pimentel2021}.

\section{Threats To Validity}
\label{Section:threats}
We limited our survey participants to a single institute, which impacts the generalizability of our findings. Additionally, while we focused on individuals actively engaged in scientific research, our convenience and snowball sampling methods may have resulted in the exclusion of suitable candidates. Moreover, those who chose to respond to our survey might already prioritize code readability. To address potential subjectivity in our analysis of free-text responses, three authors independently categorized the responses and then reached a consensus.
Finally, our research scope may not capture all aspects of code readability practices and challenges, posing a threat to the internal validity of our study. However, as this is an exploratory study, our goal is to identify preliminary trends and patterns that can guide future research in this area.

\section{Conclusion \& Future Work}
\label{Section:conclusion}
Readable and understandable code is vital for effective collaboration and reproducibility. Despite this importance, there is a lack of research on this topic in scientific programming.

This preliminary study begins to address this gap by surveying 57 research scientists across various disciplines. Our findings show that most scientists lack formal training in writing readable code, and while they recognize the importance of code readability for reproducibility, they face challenges with code comprehension. These challenges include insufficient documentation and poor identifier names, such as short and ambiguous names. Further, while a fair number never use automated code quality tools, those who do primarily rely on large language models, particularly ChatGPT. Our future work will investigate the relationship between identifier names and program behavior in scientific contexts, including the evolution of names over time and their impact on collaboration.

\bibliographystyle{ieeetr}
\bibliography{main}
\end{document}